\documentclass[12pt,preprint]{aastex}

\newcommand{\ngc}{NGC\,6231}
\newcommand{\cpd}{CPD\,$-$41$^{\circ}$7733}

\newcommand{\lam}{$\lambda$}
\newcommand{\hea}{\ion{He}{1}}
\newcommand{\heb}{\ion{He}{2}}

\newcommand{\mgb}{\ion{Mg}{2}}
\newcommand{\nb}{\ion{N}{2}}
\newcommand{\nc}{\ion{N}{3}}
\newcommand{\ob}{\ion{O}{2}}
\newcommand{\oc}{\ion{O}{3}}
\newcommand{\cb}{\ion{C}{2}}
\newcommand{\cc}{\ion{C}{3}}
\newcommand{\feb}{\ion{Fe}{2}}
\newcommand{\sic}{\ion{Si}{3}}
\newcommand{\sid}{\ion{Si}{4}}
\newcommand{\sd}{\ion{S}{4}}

\newcommand{\kms}{km\,s$^{-1}$}
\newcommand{\cnts}{cnt\,s$^{-1}$}
\newcommand{\ergs}{erg\,s$^{-1}$}

\newcommand{\rsol}{R$_{\sun}$}
\newcommand{\msol}{M$_{\sun}$}

\newcommand{\xmm}{{\sc xmm}\emph{-Newton}}

\newcommand{\epic}{{\sc epic}}
\newcommand{\mos}{{\sc mos}}
\newcommand{\pn}{pn}

\newcommand{\epicmos}{{\sc epic mos}}
\newcommand{\epicpn}{{\sc epic} pn}

\newcommand{\xspec}{{\sc xspec}}
\newcommand{\mek}{{\tt mekal}}

\newcommand{\arf}{{\it arf}}
\newcommand{\rmf}{{\it rmf}}

\defcitealias{HCB74}{HCB74}
\defcitealias{LM83}{LM83}
\defcitealias{PHYB90}{PHYB90}
\defcitealias{SL01}{SL01}

\slugcomment{}

\shorttitle{The O-type  binary CPD\,$-$41\degr7733}
\shortauthors{H. Sana et al.}

\begin{document}


\title{Constraining the fundamental parameters of the O-type binary CPD\,$-$41\degr7733}


\author{H. Sana\altaffilmark{1}, G. Rauw\altaffilmark{2} and E. Gosset\altaffilmark{2}}
\affil{Institut d'Astrophysique et de G\'eophysique, University of Li\`ege, All\'ee du 6 Ao\^ut 17, B\^at. B5c, B-4000 Li\`ege, Belgium}
\email{hsana@eso.org}


\altaffiltext{1}{Present address: European Southern Observatory, Casilla 19001, Santiago 19, Chile}
\altaffiltext{2}{Research Associate FNRS (Belgium)}


\begin{abstract}
Using a  set of  high-resolution spectra, we studied the physical and orbital properties of the O-type binary \cpd, located in the core of \ngc. We report the unambiguous detection of the secondary spectral signature and we derive the first SB2 orbital solution of the system. The period is $5.6815\pm0.0015$~d and the orbit has no significant eccentricity. 
\cpd\ probably consists of stars of spectral types O8.5 and B3. As for other objects in the cluster, we observe discrepant luminosity classifications while using spectroscopic or brightness criteria. Still, the present analysis suggests that both components display physical parameters close to those of typical  O8.5 and B3 dwarfs.
We also analyze the X-ray light curves and spectra obtained during six 30~ks \xmm\ pointings spread over the 5.7~d period. We find no significant variability between the different pointings, nor within the individual observations. The \cpd\ X-ray spectrum is well reproduced by a three-temperature thermal \mek\ model with temperatures of 0.3, 0.8 and 2.4~keV.  No X-ray overluminosity, resulting e.g. from a possible wind interaction, is observed. The emission of \cpd\ is thus very representative of typical O-type star  X-ray emission.

\end{abstract}


\keywords{
     stars: individual: CPD\,$-$41\degr7733 --
     stars: binaries: close --
     stars: binaries: spectroscopic --
     stars: early-type --
     X-rays: individuals: CPD\,$-$41\degr7733 --
     X-rays: star}



\section{Introduction} \label{sect: intro}

Early-type stars of spectral type O, and their evolved descendants, the Wolf-Rayet stars, are among the hottest and brightest objects in the Galaxy. Through their large radiative and kinetic energy input, they have a strong impact on their surroundings and their study is relevant for a number of galactic and extra-galactic issues. However, despite their importance, a proper mapping of their physical parameters is still lacking. This mainly results from the limited amount of observational constraints available so far. Indeed, only about 30 double-lined spectroscopic binaries (SB2) with O-type components have definite orbits \citep[e.g.][]{Gie03}. In this context, the present paper is devoted to a rather poorly known object, \cpd, which lies in the core of the young open cluster \object[NGC 6231]{NGC~6231}.

 \object[CPD-41 7733]{\cpd} (\object[CD-41 11037]{CD$-$41\degr11037}, \object[Braes 941]{Braes\,941}, \object[NGC 6231 323]{Se\,323}\footnote{ The SIMBAD astronomical data base reports  the \citeauthor{Se68a} number of \cpd\ to be 297. According to \citet{RCB97}, this number belongs to an extension of the \citet{Se68a} catalogue -- that only listed 295 stars -- included in WEBDA data base \citet{Me88, Me92}. However, \citet{RCB97} quoted a magnitude of 12.5 for the star \object[NGC 6231 297]{Se\,297}. This does obviously not correspond to \cpd. From the positions on the CCD, the magnitudes and the spectral types of the objects listed in their catalogue, we found that the \citeauthor{Se68a} number reported by \citet{RCB97} for \cpd\ should rather be  \object[NGC 6231 323]{Se\,323}.}) is a bright O-type star ($V=7.875$, \citealt{SBL98}; O9~III, \citealt{Wal73}).
It was first suspected to be a binary by \citet{Str44} because of the large difference between the star velocity and the cluster mean velocity. Since then, three SB1 orbital solutions have been published, but yielded discrepant orbital parameters (see \S\ref{ssect: lit}). Clearly, a new study, based on high quality data, was needed to search for the companion signature and to definitely upgrade the orbital (\S\ref{sect: spectro}) and physical (\S\ref{sect: physic}) parameters of this system. In \S\ref{sect: xmm}, we also investigate the \cpd\ X-ray properties  using a recent 180~ksec \xmm\ observation of the \ngc\ cluster. Finally, we give, in \S\ref{sect: ccl}, a summary of the main results of this work.

\section{Optical and X-ray observations} \label{sect: obs}

The present work is based on over 30 high-resolution high S/N spectra of \cpd\ obtained at the European Southern Observatory (ESO, La Silla, Chile) and spanning a time interval of six years. Table~\ref{tab: opt_diary} provides the Julian dates at mid-exposure time, the instrumental setups and the mean radial velocity (RV) at each epoch. Prior to averaging, the individual line RVs  were referred to a zero-systemic velocity frame using the systemic velocities quoted in Table \ref{tab: orbit_ne} (the considered lines are listed in Table \ref{tab: orbit_ne}). For some epochs, no reliable measurement could be obtained for the secondary, either because of the low S/N ratio or because of the important blend between primary and secondary lines. In those cases, the  corresponding column is left empty. Instrumental setups and reduction techniques have been described in e.g.\ \citet{SGR06_219} and will not be repeated here.
 
\begin{figure*}
   \centering
   \includegraphics[width=.9\textwidth]{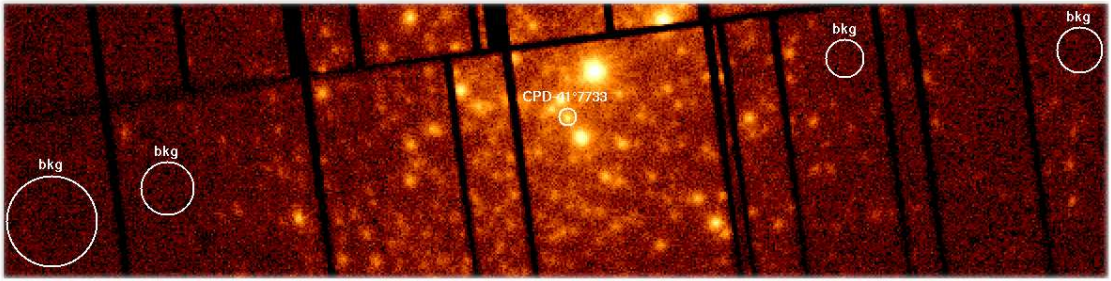}
      \figcaption{Combined \epicpn\ image in the range 0.5-10.0\,keV. Adopted source and \pn\ background (bkg) extraction regions are indicated. North is up, East to the left.
        \label{fig: pn}       }
        
   \end{figure*}

In parallel to the optical monitoring, six 30~ks X-ray exposures of the NGC~6231 cluster were performed by \xmm\ in September 2001. The campaign was described in \citet{SGR06} and we restrain the description here to additional elements specific to \cpd. For each of the six pointings and for the three \epic\ instruments, we extracted broad band X-ray light curves and spectra using the SAS task {\it evselect}. The circular extraction region was centered on the source position and, due to the presence of three bright X-ray neighbors (\object[HD 152249]{HD\,152249}, \object[HD326329]{HD\,326329} and \object[Cl* NGC 6231 SBL 324]{NGC 6231 SBL 324}),  was limited to a radius of 11\farcs5 (Fig.\ \ref{fig: pn}).  The different backgrounds were estimated from the very few source-free regions in the field of view  \citep[for more details, see][]{SRN06}.  
Finally, we also extracted the merged spectra for each instrument, thus combining the six observations of \cpd. For this purpose, we built the corresponding  \arf\ files using the SAS task {\it arfgen}, and we adopted the \rmf\ files provided by the SOC and adapted to the particular position of \cpd\ on the detectors. 
 Julian Date (JD) at mid-exposure, effective exposure times and background-corrected count rates as obtained for the different instruments are quoted in Table \ref{tab: journal}.

\section{\cpd\  orbital solution} \label{sect: spectro}
\subsection{Optical spectrum}\label{ssect: spectrum}

\begin{figure*}
   \centering
   \includegraphics[width=.9\textwidth]{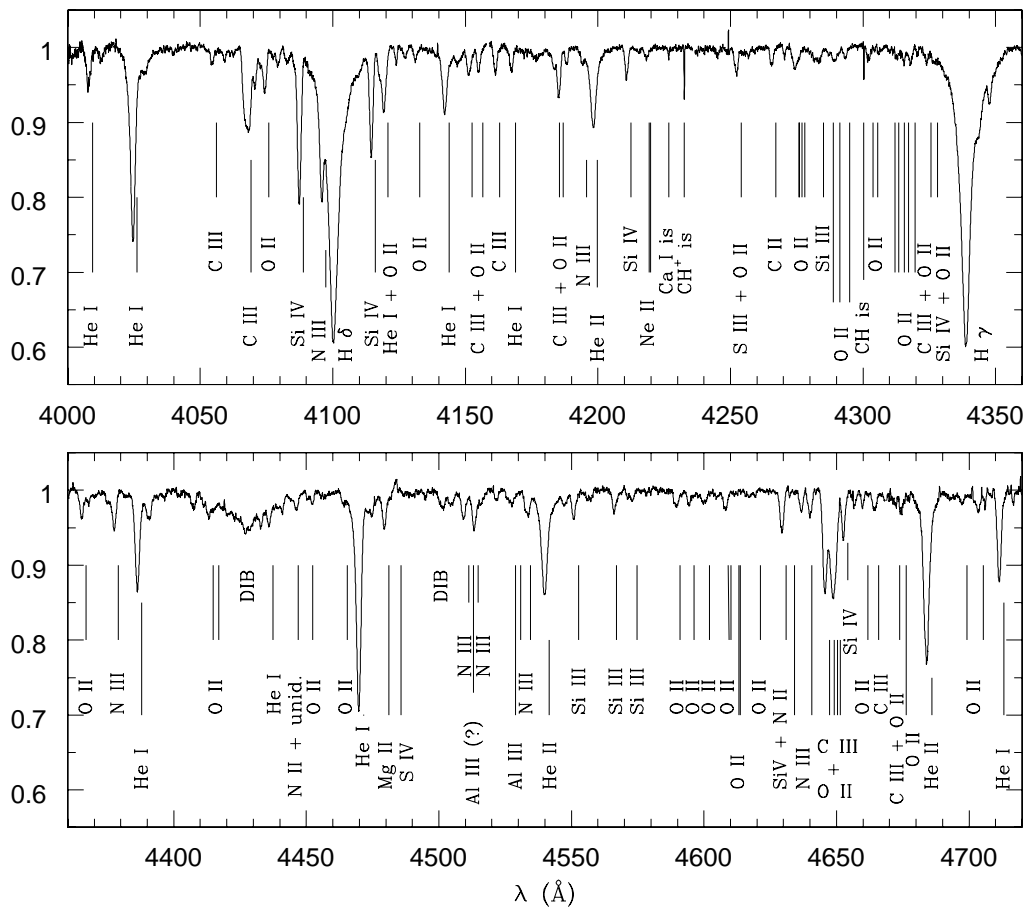}
   \figcaption{Blue spectrum of \cpd\ as obtained on HJD = 2\,453\,135.718 ($\phi=0.750$). The identifications of the main lines have been indicated. The \hea\ lines display a clear SB2 signature, with the blue-shifted primary and the red-shifted secondary components. The identification ticks refer to the rest wavelengths. \label{fig: spec} } 
  
\end{figure*}

\begin{figure}
   \centering
   \includegraphics[width=.4\textwidth]{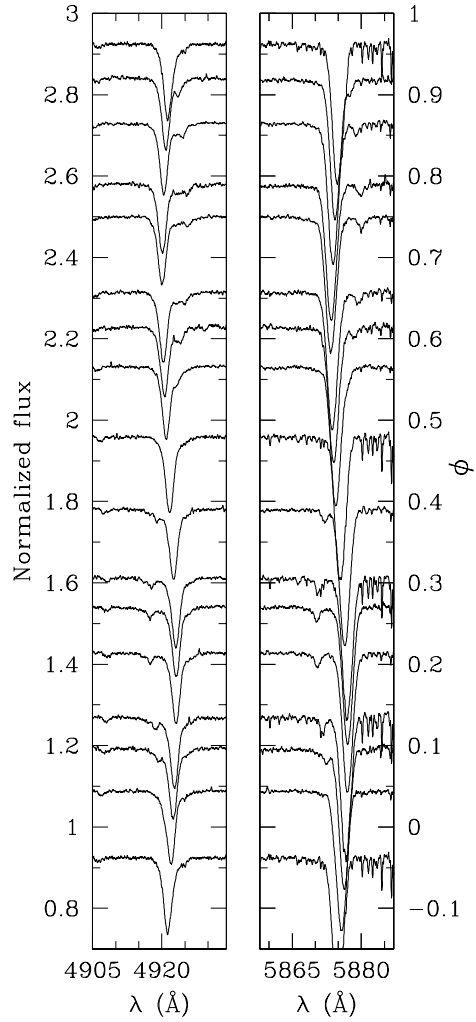}
   \figcaption{\hea\,\lam4922 (left) and \hea\,\lam5876 (right) lines at selected phases. The spectra were shifted along the vertical axis according to the value of their phase (right-hand scale). The secondary spectral signature is clearly identified in all but the blended spectra.\label{fig: doppler}  } 
   
\end{figure}

The spectrum of \cpd\ (Fig.\ \ref{fig: spec}) is clearly dominated by the \ion{H}{1} Balmer, \hea\ and \heb\ absorption lines. Numerous metallic lines (C, N, O, Si and Mg ions) can also be identified. The Balmer lines and all the \hea\ lines present a clear SB2 signature with the primary lines being several times stronger than the associated secondary lines (see e.g. Fig.\ \ref{fig: doppler}). In the very high S/N ratio spectra, the secondary signature might be seen in a few metallic lines (e.g. \cb\,\lam4267, \sic\,\lam\lam4552-68-74) but this remains at the very limit of detection. Two emission lines, \sd\,\lam4486 and \cc\,\lam5696, are further detected in the \cpd\ spectrum; both are associated with the primary component.

Line positions and equivalent widths ($W_\lambda$)  were measured by adjusting Gaussian curves to the studied profiles. Effective wavelengths for O-stars from \citet{CLL77} below 4800~\AA\ and from \citet{Und94} above were adopted to compute the radial velocities (RVs). For the metallic lines that are not listed in these latter works, we used the rest wavelengths from \citet{Moore59}.

\subsection{Orbital solution }\label{ssect: orbit}

To search for the orbital period $P$ of the system, we applied both the method of \citet{LK65} and the Fourier analysis of \citet[][ see also \citealt{GRR01} for comments]{HMM85} on the RV sets associated with the \hea\,\lam4471, \heb\,\lam4686, \oc\,\lam5592 and \hea\,\lam5876 lines. The obtained periods are all in excellent agreement with an average value of 5.68156~d. Estimating the period uncertainty on the basis of one tenth of the periodogram peak width, we found  $\sigma_P = 1.5$ to $1.8\,10^{-3}$\,d according to the data set considered.

We then computed orbital solutions using individual RV sets associated with the different absorption lines listed in Table \ref{tab: orbit_ne}. 
 For SB2 solutions, the RV equations were transformed in order to symmetrically propagate the errors on the parameters associated with each of the components \citep{SGR06_219}. We adopted a period value and a relative primary to secondary weight ratio ($s_y/s_x)$ that yield  the lowest $\chi^2$.
 We tested both circular and eccentric orbits but found that the latter did not improve the quality of the fit.  
Table \ref{tab: orbit_ne} lists the orbital solutions obtained for the data sets associated with different lines. They are all in excellent agreement with each other. As a final step, we computed the averaged RVs of all the primary lines  and of the SB2 \hea\ lines. For this, we shifted the individual RVs to a common velocity frame, taking into account the different systemic velocities deduced from the individual orbital solutions quoted in Table \ref{tab: orbit_ne}. The orbital solutions obtained using either the averaged primary line RVs only or the averaged \hea\ SB2 RVs are given in Table \ref{tab: orbit} (respectively  labeled 'Prim.' or '\hea\ lines').  The quoted period uncertainties  correspond to the 1-$\sigma$ confidence interval\footnote{ The confidence intervals on the period value were computed by varying the period (all other parameters being kept constant at  their best-fit value) until the minimum $\chi^2$ value of the fit is increased by a value $\Delta\chi^2$. The latter value was chosen to  correspond to the 68.27\% confidence level and depends on the number of degrees of freedom of the fit. }. We emphasize however that the uncertainty on the periodicity of the phenomenon (which is independent of any model consideration) is rather related to the peak width in the periodogram as stated at the beginning of this section. The other uncertainties (always quoted as one standard deviation) were computed by error propagation from the least-square  fit.
Finally, we adopted the apparent systemic velocities of both components as the weighted means of the systemic velocities  in the different \hea-line solutions of Table \ref{tab: orbit_ne}:  $\overline{\gamma_1}=-21.5\pm0.2$\,\kms\ and  $\overline{\gamma_2}=-23.1\pm0.2$\,\kms.

\subsection{Towards a global solution including published and new data } \label{ssect: lit}

As mentioned earlier, three SB1 orbital solutions have previously been published. Based on a set of 16 observations spread over 8.1 days plus four observations from \citet[][\citetalias{PHYB90} hereafter]{PHYB90} obtained about 800 days earlier, \citet[\citetalias{HCB74} hereafter]{HCB74} derived the first SB1 orbital solution with a period $P=5.64\pm0.01$\,d, a small eccentricity $e=0.04\pm0.03$ and an amplitude for the  primary radial velocity curve  of $K_1=89.0\pm2.7$\,\kms. With two more observations  and including radial velocities from \citet{Str44} and \citet{HCB74} (but not those of \citealt{PHYB90}, only published much latter), \citet[\citetalias{LM83} hereafter]{LM83} derived a new orbital solution. They found a larger period ($P=5.74973\pm3\,10^{-5}$\,d) and eccentricity ($e=0.14\pm0.015$) while the amplitude of the radial velocity curve remained mostly unchanged ($K_1=84\pm2$\,\kms). Finally, \cpd\ was observed once by IUE. Including this additional measurement, \citet[ \citetalias{SL01} hereafter]{SL01} published a slightly modified orbit with $P=5.749809\pm3.0\,10^{-5}$\,d, $e=0.045\pm0.021$ and $K_1=83.9\pm2.1$\,\kms. They stated however that they could not include the \citetalias{PHYB90} data in their fit. 

Though the present data set yields orbital parameters of the same order of magnitude than the ones obtained from previous works, fundamental contradictions exist between the different studies. Our final period $P=5.681504$\,d is significantly different from the previous determinations and we report a circular orbit while, for example, \citetalias{LM83}  quoted  an eccentricity up to $0.14\pm0.015$. Our derived semi-amplitude of the primary RV curve $K_1$ is systematically higher than the previously published values, with differences up to 10\,\kms\ compared to \citetalias{LM83} and \citetalias{SL01}.  It was thus crucial to check the previous orbital solutions using the different RV sets adopted by these authors. 
Doing this, we noted that the second measurement listed by \citetalias{PHYB90}, at $RV = -109$\,\kms, is clearly not fitted by the \citetalias{HCB74} solution though it was included in the adjustment. Actually its location in the phase diagram is shifted by about 0.2 in phase compared to the figure displayed by \citetalias{HCB74}. This  probably results from a typographic error in the published Julian date, which  explains why \citetalias{SL01} could not include \citetalias{PHYB90} observations in their combined solution.

Though we have been able to reproduce the results of  \citetalias{HCB74} and  \citetalias{SL01} using their respective data sets, we could not reproduce the results proposed by \citetalias{LM83}. Both the period value and, more particularly, the eccentricity derived from their data differ significantly.  In all our computations, whatever the data set used, the maximum eccentricity that we found was $e=0.07$, with a null eccentricity well within a 2-$\sigma$ interval. We thus definitively rule out the large eccentricity quoted by  \citetalias{LM83}. Furthermore, both the \citetalias{LM83} and \citetalias{SL01} data sets actually carried a strong ambiguity on the period, with more than ten strong aliases between 5.6 and 5.8 days. Clearly, their period determination was far less constrained than what their error bars suggested.

\begin{figure}
   \centering
   \includegraphics[width=.9\columnwidth]{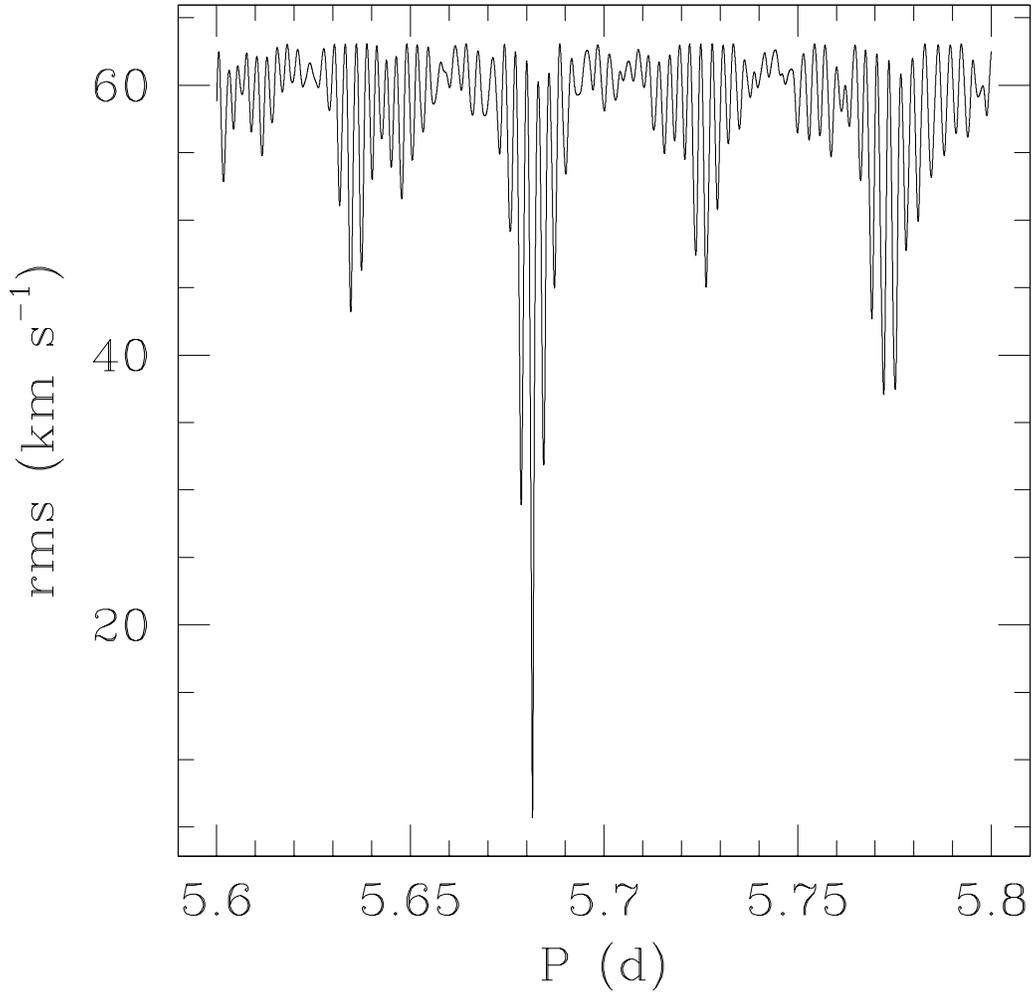}
      \figcaption{Root-mean square (r.m.s.) residuals of the best circular orbital solutions as a function of the adopted period. The adjustments were performed using the different RV sets presented in Fig.\ \ref{fig: rv_lit}, excluding the point of \citetalias{PHYB90} at $RV=-109$\,\kms. 
            \label{fig: alias}     }
      
   \end{figure}

\begin{figure}
   \centering
   \includegraphics[width=.9\columnwidth]{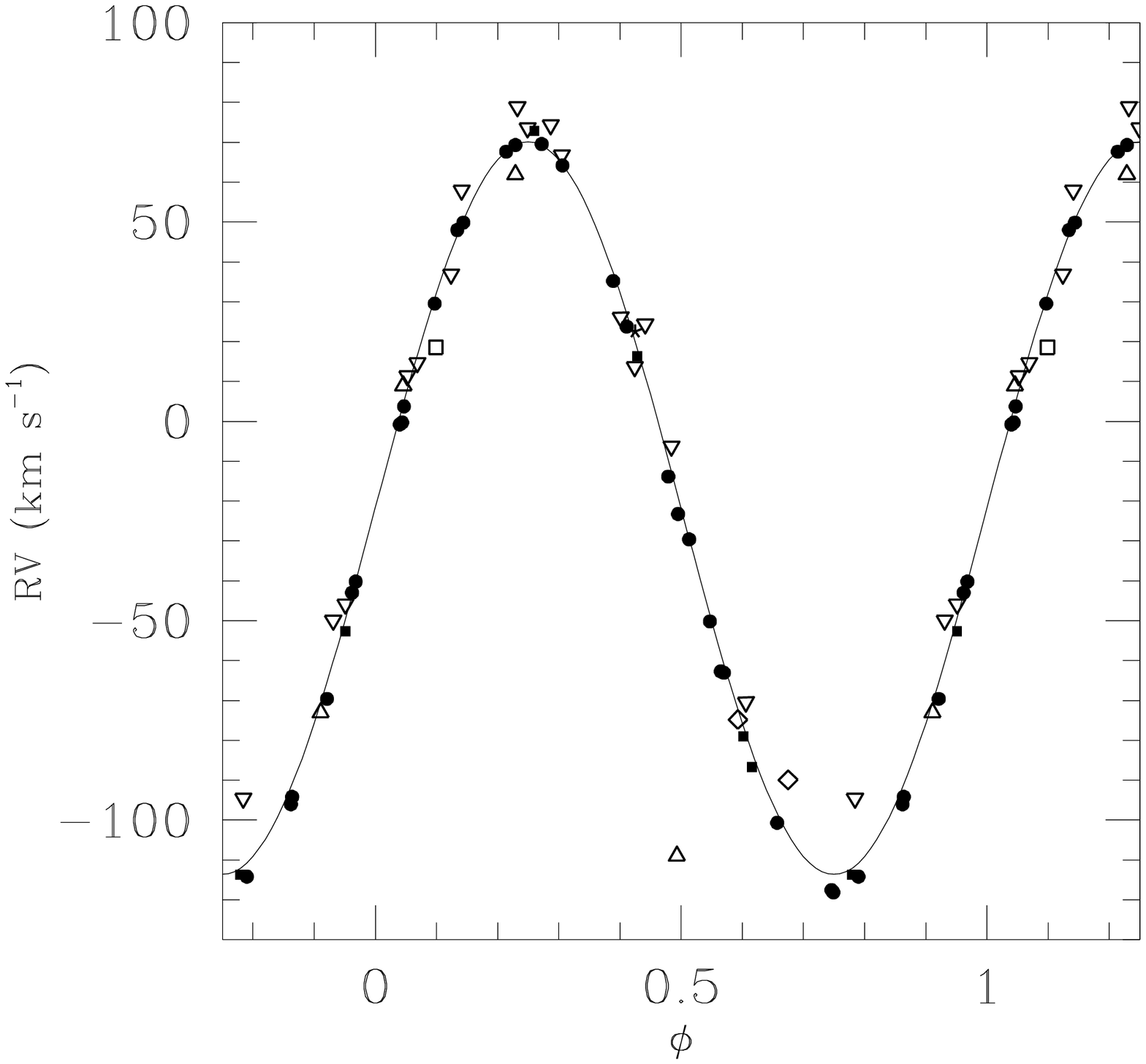}
      \figcaption{Combined  RV curve computed using literature data together with the measurements of Table \ref{tab: opt_diary}. Different symbols refer to different data sets. Open symbols are for previous observations : \citealt{Str44}, asterisk;  \citetalias{PHYB90}, upward triangles;  \citetalias{HCB74}, downward triangles; \citetalias{LM83}, diamonds;  \citetalias{SL01}, square. Filled symbols indicate new RV points from this work: CES, squares; FEROS, circles. Note that  the \citetalias{PHYB90} point at $RV=-109$ \kms, near $\phi\approx0.5$, was excluded for the period search and orbital solution determination.
        \label{fig: rv_lit}      }
 
   \end{figure}

Finally, we combined all the available primary RV measurements (except the \citetalias{PHYB90} point at $RV=-109$\,\kms) to derive a combined orbital solution. For the RVs from the present work, we used the averaged RVs reported in Table \ref{tab: opt_diary} to which we added the adopted primary  systemic velocity $\overline{\gamma_1}=-21.5$\,\kms. A period search using both the Lafler \& Kinman and the Fourier analysis techniques (see \S\ref{ssect: orbit}) yield values close to 5.6815 d with an uncertainty of $\sigma_P=2.4\ 10^{-4}$ d. Again no significant improvement of the quality of the fit is obtained assuming an eccentric orbit. Fig.\ \ref{fig: alias} shows the values of the fit r.m.s.\ for periods ranging from 5.6 to 5.8 d. Clearly, the present analysis allows us to solve the ambiguity about the period that plagued earlier investigations. The best-fit orbital parameters are listed in Table \ref{tab: orbit} (labeled 'Lit.') and the RV curve is displayed in Fig.\ \ref{fig: rv_lit}. The corresponding heliocentric systemic velocity is $-21.7\pm0.1$~\kms.

\section{\cpd\ physical parameters} \label{sect: physic}
\subsection{Spectral types and luminosity classes } \label{ssect: spt}
\subsubsection{Primary component}

The spectral signature of the primary component is easily discernible in the spectrum of \cpd. Adopting the classification criteria from \citet{Con73_teff} as adapted to late O-stars by \citet{Mat88}, we obtained a mean $ \log W' (\frac{ W_{\lambda4471} }{ W_{\lambda4542} })=0.22\pm0.05$ which corresponds to a spectral type O8.5, with spectral type O8 within $1\,\sigma$.
To determine the luminosity class, we adopted the criterion from \citet{CA71}. We obtain $\log W''(\frac{W_{\lambda4089}}{W_{\lambda4144}})=0.27\pm0.04$, which leads to a giant luminosity class. 
We also measured $\log W''' = \log (W_{\lambda4388}) + \log(W_{\lambda4686})=5.23\pm0.04$. According to \citet{Mat88}, this rules out a supergiant class and points towards a giant classification, unless the ratio $l_1=\frac{L_1}{L_\mathrm{tot}}<0.67$. The optical brightness of both components will be discussed in \S\ref{sect: brightness}. 

\subsubsection{Secondary component}
The main spectral signatures of the secondary star in \cpd\ are the Balmer and \hea\ lines. The absence of the \heb\ lines, as well as of the \oc\,\lam5592 line, at the positions predicted by the orbital solution definitively excludes an O spectral type and, at our detection threshold, most probably indicates a spectral subtype later than B0.7 \citep{WF90}. To refine our classification, we searched for the presence of secondary  metallic lines. This led us to consider low intensity lines, with equivalent widths (relative to the composite spectrum continuum) down to $\sim$0.01\AA. We therefore focused on the six FEROS spectra obtained at the ESO/MPG 2.2m telescope, that exhibit S/N ratios above 200.  Mean primary and secondary equivalent widths as measured on these spectra are reported in Table \ref{tab: lines_list}. Some of the secondary lines are  clearly deblended from neighboring lines, and thus securely measured, in only one or two of these six spectra. These are marked with a colon (:) and should be considered as a rough indication of the line strength only.

The absence of \heb\,\lam4686, \cc\,\lam4650, \nc\,\lam4097 and \sid\,\lam\lam4089, 4116 lines in the secondary spectrum, combined with the presence of a comparatively strong \cb\,\lam4267 line, point towards a spectral type later than B2 while the lack of \feb\ lines corresponds to a spectral type earlier than B5. 
The usual luminosity criterion in this spectral range is \sic\,\lam4552/\hea\,\lam4388. From our measurements, we obtained a ratio of about one to ten; this result definitively excludes a supergiant luminosity class, which is in agreement with the low intensity of the secondary \ob\ spectrum. Similarly, this ratio seems to indicate a main sequence luminosity class rather than a giant class.

\subsection{Optical brightness ratio } \label{sect: brightness}

We estimated the optical brightness ratio based on the dilution of the primary lines. We compared their mean $W_\lambda$\ with typical (averaged) $W_\lambda$\ of O8.5\,III stars \citep{CA71, Con73_nlte, Mat88, Mat89}. Based on the \hea\,\lam\lam4026, 4388, 4471 and \heb\,\lam\lam4542, 4686 lines, we obtained an averaged brightness ratio corresponding to $l_1=0.87\pm0.10$. We then compared the secondary intrinsic $W_\lambda$\ with typical $W_\lambda$\ for B stars \citep{Did82} using a dilution factor $l_2=0.13$. The intensity of the detected lines in the secondary spectra is in rough agreement with a B3 III-V classification but suggests a slightly larger luminosity ratio in favor of the primary component.
Adopting $l_1 = 0.87 \pm 0.10$ yields $\log W'''_\mathrm{prim}  \approx 5.29 \pm 0.06$, which corresponds to a giant luminosity class for the primary though we note that the \heb\,\lam4686 line is  stronger than in typical O8.5 giants. 
Should we assume that the primary is a main sequence star, the comparison of the observed $W_\lambda$\ with typical $W_\lambda$\ for O8.5~V stars yields $l_1=0.74\pm0.08$. This yields $\log W'''_\mathrm{prim}  \approx 5.36 \pm 0.06$ thus in contradiction with the V luminosity class hypothesis. Clearly the spectroscopic classification criteria points towards the primary being a giant.

\subsection{Luminosities and stellar radii}

Typical absolute visual magnitudes  for O8.5 stars are $M_\mathrm{V}=-5.2$ and $M_\mathrm{V}=-4.5$/$-4.4$ for class III and V respectively \citep{HM84, HP89}. The absolute visual magnitudes of B3 stars are $M_\mathrm{V}=-3.0$ and $-1.6$ for giants and main-sequence stars respectively \citep{SK82, HM84}. To compute the visual magnitudes of the \cpd\ components, we adopted the cluster distance modulus $DM=11.07\pm 0.04$ as obtained from the average of the different photometric results since the 1990's \citep{SGR06} and we used $V=7.875$, $R=3.3$ and $(B-V)=0.158$ from \citet{SBL98}. 
Using $l_1=0.87\pm 0.10$ as derived in the previous paragraph in the case of a giant primary and assuming that the secondary  is also a giant, we obtained $M_\mathrm{V}=-4.67~\pm~0.15$ for the total magnitude of the system, and 
 $M_\mathrm{V,1}=-4.52~\pm~0.20$ and  $M_\mathrm{V,2}=-2.46~\pm~1.12$ for the primary and secondary respectively. Under these hypotheses, the primary is clearly fainter than typical O8.5 giants. 
Adopting the effective temperature calibration and the bolometric correction of \citet{HM84}, we constrained the radii to values of $R_1=8.8\pm1.0$ \rsol\ and $R_2=6.3\pm3.9$~\rsol. Again, the stars are too small compared to typical giants ($R^\mathrm{O8.5~III}\approx 13$~\rsol\ and $R^\mathrm{B3~III}\approx 9$~\rsol).

Therefore, if we assume that the \cpd\ components are indeed main sequence objects, the same reasoning with $l_1=0.74\,\pm\,0.08$ yields  $M_\mathrm{V}=-4.64\,\pm\,0.13$, $M_\mathrm{V,1}=-4.31\,\pm\,0.17$ and  $M_\mathrm{V,2}=-3.18\,\pm\,0.37$, $\log(\frac{L^\mathrm{bol}_1}{L_{\sun}})=4.98\,\pm\,0.08$, $\log(\frac{L^\mathrm{bol}_2}{L_{\sun}})=3.89\,\pm\,0.17$  and therefore $R_1 = 8.4\,\pm\,1.0$~\rsol\ and $R_2 = 8.4\,\pm\,1.9$~\rsol. Corresponding typical radii are, in this case, of 9.0~\rsol\ and 4.2~\rsol\ for O8.5 and B3 main sequence stars. We now observe a better agreement for the primary, but the secondary parameters are at odds with those of typical B3 dwarfs.

The best agreement between the deduced  and typical parameters is obtained assuming that \cpd\ harbors an O8.5~V and a B3~III component. In the latter case, we derived  $M_\mathrm{V,1}=-4.31\pm0.17$, $M_\mathrm{V,2}=-3.18\pm0.37$, $R_1 = 8.4 \pm 1.0$~\rsol\ and $R_2 = 8.7\pm1.7$~\rsol. This option is however rather unlikely.  We finally note that a larger value for $l_1$ would significantly decrease the obtained secondary radius without affecting much $R_1$. Such a ratio largely in favor of the primary ($l_1>0.9$) is indeed suggested by the typical O8.5 and B3 luminosities.

\begin{figure}
\centering
\includegraphics[width=.9\columnwidth]{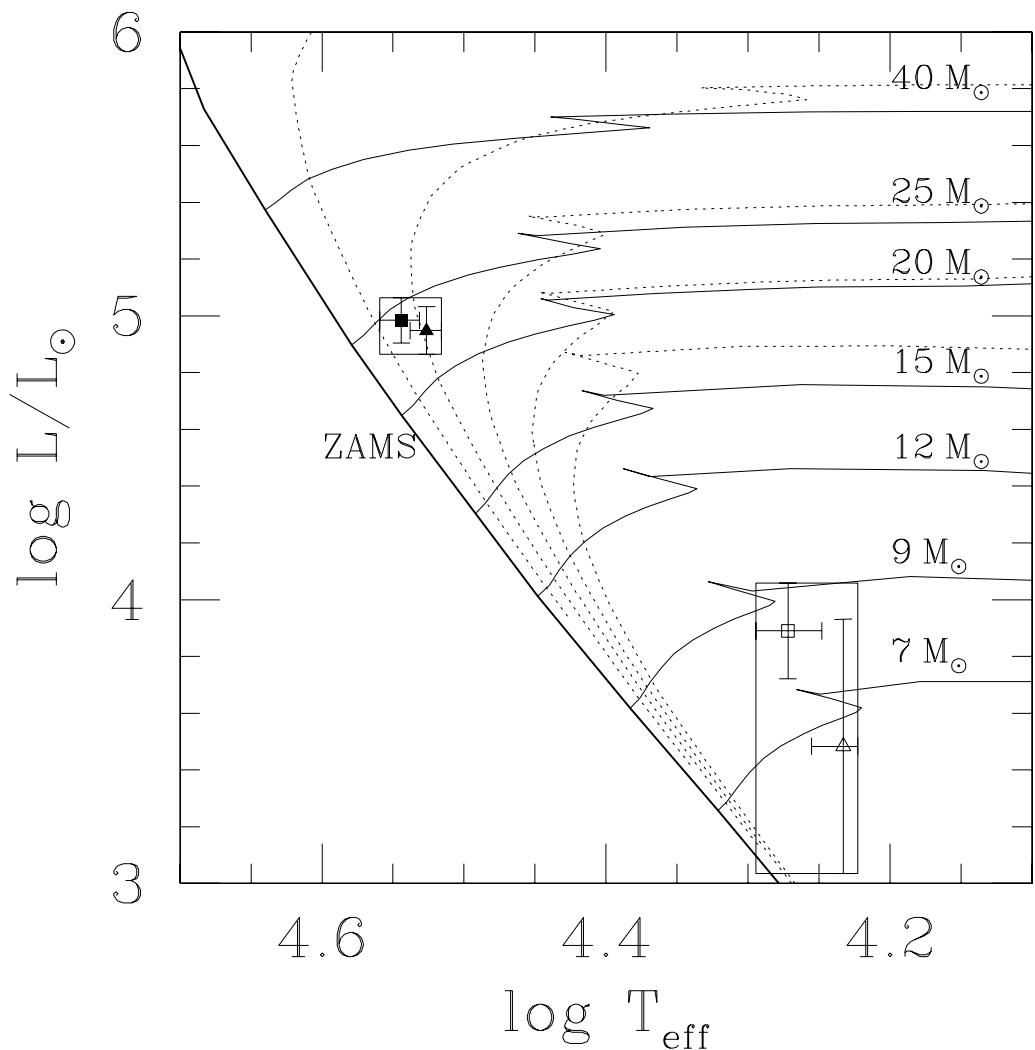}
\figcaption{Location of the \cpd\ primary (filled symbols) and secondary (open symbols) components in the H-R diagram. The triangles (resp. squares) indicate an adopted giant (resp. main sequence) luminosity class. The evolutionary tracks from \citet{SSM92} are shown (plain lines) as well as the isochrones (dotted lines) computed for ages ranging from 2 to 10 Myr with a step of 2 Myr. The boxes present the ranges of parameter values obtained assuming different luminosity classes for the components.\label{fig: HR}}

\end{figure}

In summary, though the spectral criteria rather indicate a giant luminosity class for the primary, the estimated values for its magnitude and radius are more consistent with the primary being a main sequence star. This situation is reminiscent of the case of \object[CPD-41 7742]{CPD$-$41\degr7742}, an eclipsing SB2 binary in \ngc\ for which the spectroscopic criteria clearly indicated a giant class. In \citet{SHRG03}, we inferred, from the system luminosity,  much smaller radii than expected for giant stars. This was confirmed by the analysis of the system light curve \citep{SAR05} and the resulting parameters are in good agreement with the typical values expected for dwarfs of the corresponding spectral types. 
The locations of the \cpd\ components in the H-R diagram (Fig.\ \ref{fig: HR}) indicate a primary evolutionary age of 3 to 4~Myr, in good agreement with previous age estimates for the cluster. This suggests that the primary is only just leaving the main sequence and could explain why the star displays properties somewhat intermediate between those of a dwarf and a giant. In this respect, the estimate of the brightness ratio based on the dilution of the primary lines might have been biased. It is thus probable that both components display stellar radii typical of  main sequence objects.

\subsection{Masses and orbital inclination}
Typical masses for O8.5~V (resp. III) stars are about 19 (resp. 24) \msol\ \citep{Martins}.  Comparing this with the minimal values obtained in Table \ref{tab: orbit} indicates that the orbital inclination of the system is probably around 74\degr\ (resp. 63\degr). This corresponds to a secondary mass around 7 to 9~\msol, which is slightly lower than typical values of B3~III stars \citep[see e.g.][]{SK82}, but is consistent with B3 main sequence stars. 
Adopting the different values for the radii derived in the previous section, we found that \cpd\ should not display eclipses unless $i\ge65-70\degr$. Clearly this is a limiting case.
 Referring to a paper in preparation by \citeauthor{PHYB90}, \citet{HCB74} announced that the star was displaying magnitude variations larger than 0\fm15. This result was however not confirmed once the mentioned paper finally appeared as \citet{PHYB90}.  From our photometric campaign of the cluster \citep{SAR05}, the r.m.s.\ of the data set associated to \cpd\ is about 0\fm012, slightly larger than the expected noise. The peak-to-peak variations over the one month time span of our campaign is 0.05\,mag at maximum. A power spectral analysis indicates a slightly dominant peak at $\nu\approx0.084$~d$^{-1}$ in both filters, thus tentatively suggesting a period close to 12~d although this can not be the signature of eclipses. 

As a last point, we estimated the projected rotational velocity $v\sin i$ of the primary star by comparing the full widths at half maximum of several lines  with those given by model spectra computed  with an effective temperature and a gravity value corresponding to the above estimates for the primary component\footnote{ The model spectra were computed, for a grid of $v\sin i$ ranging from 50 to 300\,\kms\ with a step of 10\,\kms, using the {\sc tlusty} and {\sc synspec} codes \citep{LaH03} that use line blanketed, NLTE, plane-parallel, hydrostatic atmospheres.}. Based on the lines quoted in Table \ref{tab: orbit_ne} (with the exception of \hea\,\lam\lam4471 and 7065), we derived $v\sin i\approx 83\pm8$\,\kms. Given a probable inclination of 65-75\degr, this value is slightly larger, still compatible within the error bars, with a synchronous rotation rate (corresponding, for $R_1=8.4-8.8$\,\rsol, to $v\approx 75-78$\,\kms). No significant difference was found assuming either a dwarf or a giant luminosity class for the primary.  

\section{X-ray analysis}  \label{sect: xmm}

We first extracted broad band X-ray light curves of \cpd\ using the average count rates during each observation. The different energy bands considered are the total band [0.5-10.0 keV], a soft (S$_\mathrm{X}$) band [0.5-1.0 keV], an intermediate (M$_\mathrm{X}$) band [1.0-2.5 keV] and a hard (H$_\mathrm{X}$) band [2.5-10.0 keV]. We found no consistent variations between the three instruments. A $\chi^2$ test performed using the count rates in the different energy bands did not allow us to reject the null hypothesis of a constant count rate during all six pointings, except for the \pn\ instrument that displays a deviating point in the M$_\mathrm{X}$ band at $\phi\approx0.18$. Using the individual event lists obtained during each of the six pointings, we also searched for short term variability. We performed a time series analysis similar to the one carried out for \object[HD 152248]{HD~152248}, the central target of the field \citep{SSG04}. Again no significant variation could be consistently detected for any of the three instruments. This strongly suggests that, at our detection threshold, the X-ray emission from \cpd\ is mostly constant.

\begin{figure}
 \centering
 \includegraphics[width=.9\columnwidth]{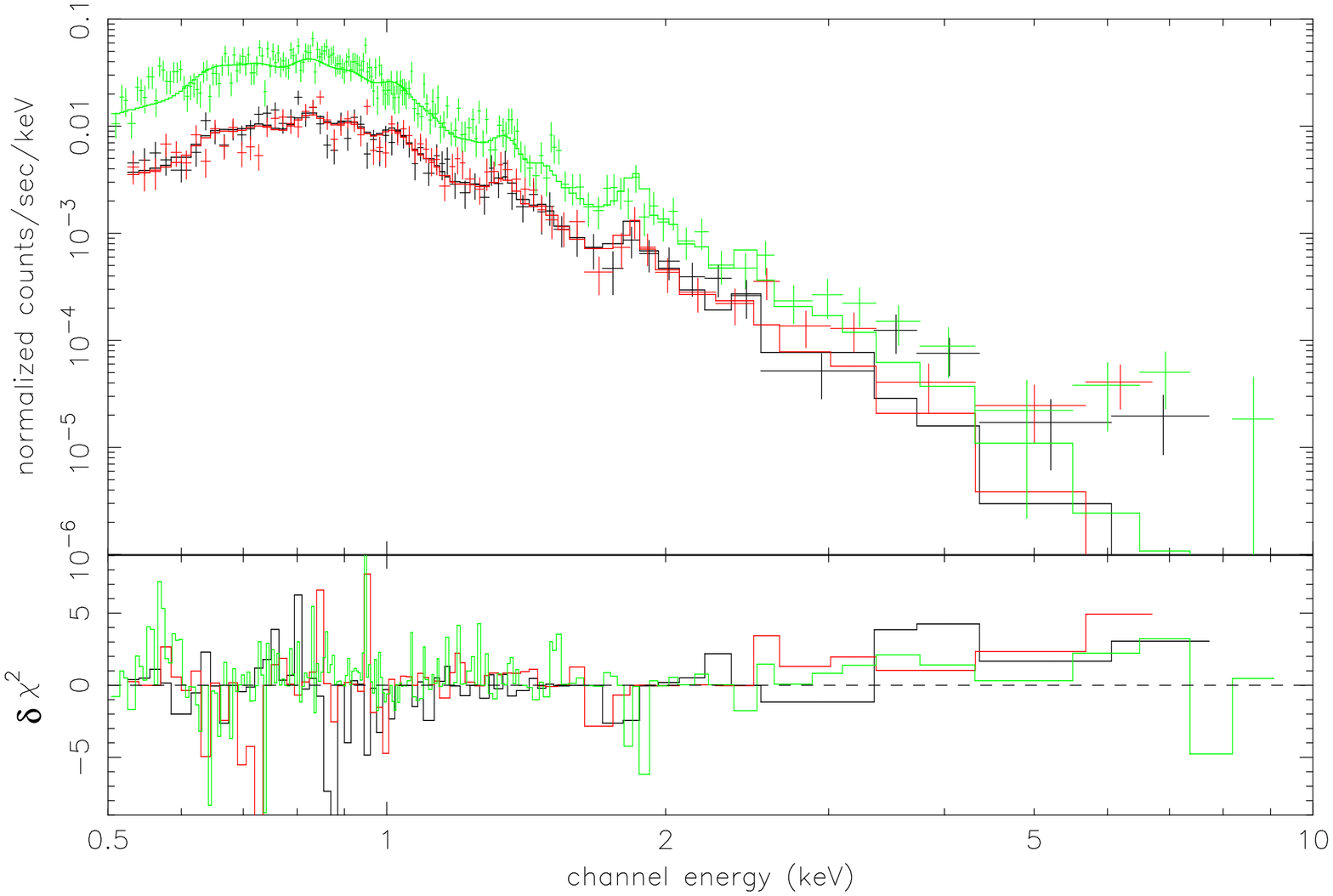}
 \includegraphics[width=.9\columnwidth]{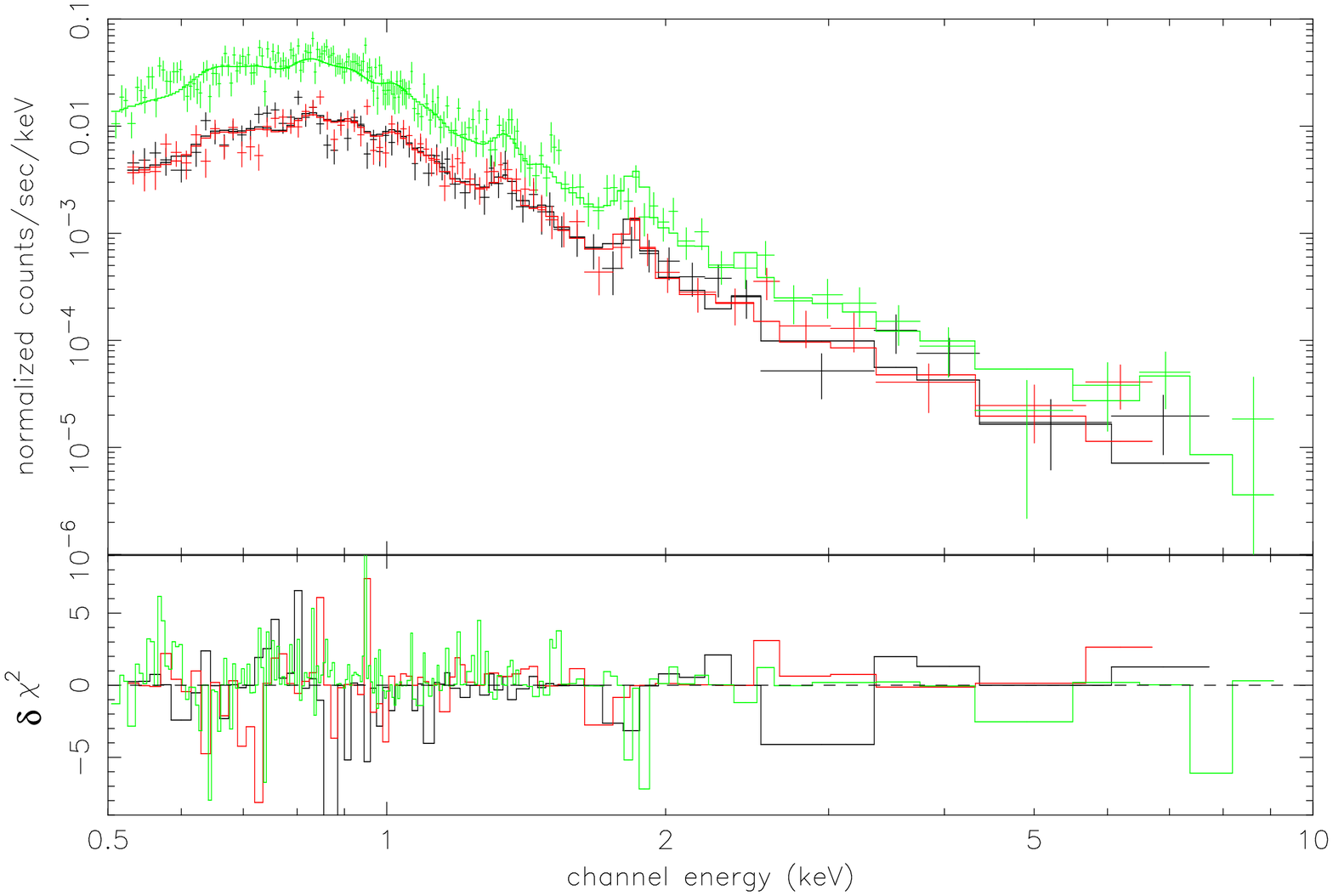}
 \figcaption{Merged \epicmos1 (black), \mos2 (red) and \pn\ (green) spectra of \cpd\ simultaneously fitted with  2-T (upper panel) and  3-T (lower panel) \mek\ models. The bottom window of each panel shows the contributions of individual bins to the $\chi^2$ of the fit. The contributions are carried over with the sign of the deviation (in the sense data minus model). The figures appear in color in the electronic version of the paper.\label{fig: 2T}}

 \end{figure}

To constrain the physical properties of the emitting plasma, we adjusted a series of optically thin thermal plasma \mek\ models \citep{MGvdO85, Ka92} to the spectra obtained during each \xmm\ observation. The  \epicmos\ and \epicpn\ spectra were adjusted simultaneously using \xspec\ v.11.2.0 \citep{arn96}. We adopted an equivalent interstellar column of neutral hydrogen of $N_\mathrm{H, ISM}=0.26\times 10^{22}$\,cm$^{-2}$, corresponding to $E(B-V)=0.447$ as obtained from \citet{SBL98}. A single temperature model was insufficient to adequately describe the observed spectra. We thus adopted two-temperature (2-T) models allowing for possible local absorption for both \mek\ components.  The absorption column associated with the lower temperature component tends to be systematically close to zero. Lower residuals and more stable solutions are obtained by fixing this column to zero. This situation is reminiscent of what was observed for HD~152248 \citep{SSG04} and CPD$-$41\degr7742 \citep{SAR05}, two other early-type binaries in the core of the \ngc\ cluster. Therefore, the actually fitted model is ({\tt wabs$_\mathrm{ISM}$ * (mekal$_1$ + wabs$_2$ * mekal$_2$)}), in which the term {\tt wabs$_\mathrm{ISM}$} was fixed to the interstellar value ($N_\mathrm{H, ISM}=0.26\times 10^{22}$\,cm$^{-2}$). Table \ref{tab: Xspec} provides the best-fit parameters ($N_\mathrm{H}$, the absorbing column; $kT$, the plasma component temperature; $norm$, the normalization factor) and the limits of the 90\% confidence intervals. Corresponding X-ray fluxes in the total energy band ($f_\mathrm{X}$) as well as in the soft ($f_\mathrm{X,S}$), intermediate ($f_\mathrm{X,M}$) and hard ($f_\mathrm{X,H}$) bands are also provided.
Finally, for each of the \epic\ instruments, we extracted the combined spectrum from the merging of the six X-ray observations. We also used a 2-T \mek\ model to fit the obtained spectra and the best parameters are given in the last line of  Table \ref{tab: Xspec}. The spectra and the best fit model are presented in Fig.\ \ref{fig: 2T}. The ISM absorption corrected fluxes ($f^{un.}$) computed in the different energy bands considered here as well as the observed X-ray luminosity in the total  0.5-10.0~keV band are given in Table~\ref{tab: xlum}.

It is clear from Fig.\ \ref{fig: 2T} that an additional, higher energy component is present in the spectrum of \cpd.  Adjusting a  three-temperature (3-T) \mek\ model reveals that the third component has a best-fit temperature close to 2.4~keV. We also adjusted a 2-T \mek\ +  power-law (PL) model. To avoid the fitting procedure of the PL component to be biased by small discrepancies at lower energy, we first adjusted the 2-T model in the 0.5-4.0 keV range. We then held these parameters fixed to their best fit values while extending the energy domain up to 10.0 keV. Again, both the 3-T and the 2-T+PL models fit the observed spectra adequately. The best-fit photon index of the PL component is about $\Gamma\approx 2.9$, but it is very poorly constrained. A PL component with $\Gamma=1.5$ fits the spectra almost equivalently well. In NGC~6231,  such a higher energy component (with a typical temperature of a few keV) is seen in four objects (HD~152248, \object[HD 326329]{HD~326329}, object[HD 152314]{HD~152314} and \cpd), which corresponds to about one fourth of the O-type stars in the cluster \citep{SRN06}. Three of them are binaries but only one, HD~152248, displays a wind-wind collision. The X-ray emission from other systems similar to \cpd\ (such as HD~152219) does not display such an additional  higher energy component. Differences between the systems with and without such a component remain unclear. 

Finally, we used the constraints on the physical parameters deduced in the previous sections to get more insight into the winds of the \cpd\ components. We assumed  an orbital inclination  of 70\degr. Mass-loss rates and terminal velocities were estimated following \citet{VdKL01}. As expected, the primary wind is overwhelmingly dominant and no ram pressure equilibrium is possible on the system axis. This suggests that the primary wind will crash into the secondary star surface. However, using the formalism of \citet{Usov},  one estimates that the  amount of X-ray emission that could be produced by such an interaction is about 10$^{29}$ \ergs, thus two orders of magnitude smaller than the intrinsic contribution of the two stellar components. As the \cpd\ orbit is circular, we further do not expect to observe an intrinsic modulation of the emitted flux resulting from a variation of the shock strength, which is thus in agreement with our observations. The X-ray emission of \cpd\ is thus very representative of the X-ray emission of normal O-type stars and, indeed, it is very well fitted by the newly  obtained canonical relation $\log(\frac{L_\mathrm{X}}{L_\mathrm{bol}})=-6.91\pm0.15$, derived for the O-type stars in \ngc\ \citep{SRN06}.

\section{Conclusions} \label{sect: ccl}
We present the results of a high-resolution spectroscopic campaign on the O-type binary \cpd. We report the first detection of the secondary spectral signature and we derive the very first SB2 orbital solution for the system. The orbital period is close to 5.681~d and the orbit is most probably circular. The new orbital elements are significantly different from those obtained in previous works. Using the same data sets as previous authors, we find that their period values were poorly constrained because of a strong aliasing that resulted from the spread of a small number of observations over a large time span. We also combined all the observations available from the literature in a joint orbital solution. The resulting orbital elements are in perfect agreement with those found using our data set alone. 

\cpd\ most probably consists of an O8.5 star plus a B3 companion. Though different criteria yield discrepant luminosity classifications, both stars probably display physical parameters close to those of typical O8.5 and B3 dwarfs. Though the orbital inclination is rather large (around 65-75\degr), our photometric campaign on \ngc\ did not reveal any eclipse. Finally, we estimated the projected rotational velocity of the primary to be around 85\,\kms, slightly larger than, but still compatible within the error bars, with the synchronous rotation rate.

We also analyze the X-ray light curves and spectra of \cpd\ obtained during the six \xmm\ pointings towards the cluster. We find no significant variability between the different pointings, nor within the individual observations. The  X-ray spectrum is well reproduced by a two-temperature thermal \mek\ model with k$T_1\approx0.3$~keV and k$T_2\approx0.8$~keV. The merged spectrum built from the combination of the six pointings shows an additional higher energy component. The latter can be described either by a \mek\ model with an energy close to 2.4 keV or by a power-law component with a photon index $\Gamma$ close to 3, though the latter value is very loosely constrained.  No X-ray overluminosity resulting from a possible wind interaction phenomenon is observed and, indeed, none is expected at our detection level. As a consequence, the emission of \cpd\ should be very representative of typical O-type star  X-ray emission.

\acknowledgments
It is a pleasure to thank the referee, Dr. D. Gies, for detailed and helpful suggestions, as well as N. Linder for running the {\sc synspec} and {\sc tlusty} codes.
The authors are greatly indebted towards the FNRS, Belgium. This work was supported by the PRODEX XMM and Integral Projects, contracts P4/05 and P5/36 `P\^ole d'Attraction Interuniversitaire' (Belgium).



{\it Facilities:}  \facility{ESO/Max Plank:2.2m (FEROS)}, \facility{ESO:1.52m (FEROS)}, \facility{ESO:CAT (CES)}, \facility{XMM (EPIC)}.






\clearpage

\begin{table}
\caption{Journal of the spectroscopic observations} \label{tab: opt_diary}
\begin{tabular}{r r r r}
\tableline
\tableline
\vspace*{-3mm} \\
HJD \hspace*{2mm} & $\phi_\mathrm{He~I}$ \hspace*{2mm}& $ \overline{RV_{1,\lambda}-\gamma_{1,\lambda}}$ &  $\overline{RV_{2,\lambda}-\gamma_{2,\lambda}}$\vspace*{1mm}\\
$-$2\,450\,000    &                     & (\kms) & (\kms)\\
\tableline
   996.684\tablenotemark{a} &  0.259 &     96.8  \hspace*{3mm} &   -261.6   \hspace*{3mm}  \\
   997.638\tablenotemark{a} &  0.427 &     40.3  \hspace*{3mm} & $-$107.4   \hspace*{3mm} \\
   998.627\tablenotemark{a} &  0.601 &  $-$55.1  \hspace*{3mm} &    149.3   \hspace*{3mm} \\
   999.634\tablenotemark{a} &  0.778 &  $-$89.8  \hspace*{3mm} &  \nodata   \hspace*{3mm} \\
  1000.607\tablenotemark{a} &  0.950 &  $-$28.7  \hspace*{3mm} &  \nodata   \hspace*{3mm} \\
  1299.827\tablenotemark{b} &  0.615 &  $-$62.3  \hspace*{3mm} &    156.6   \hspace*{3mm}  \\
  1300.818\tablenotemark{b} &  0.790 &  $-$90.3  \hspace*{3mm} &    238.2   \hspace*{3mm} \\
  1301.825\tablenotemark{b} &  0.967 &  $-$14.6  \hspace*{3mm} &  \nodata   \hspace*{3mm} \\
  1302.827\tablenotemark{b} &  0.143 &     74.1  \hspace*{3mm} & $-$190.8   \hspace*{3mm} \\
  1304.822\tablenotemark{b} &  0.494 &      1.0  \hspace*{3mm} &      3.5   \hspace*{3mm} \\
  1327.843\tablenotemark{b} &  0.546 &  $-$24.8  \hspace*{3mm} &     83.8   \hspace*{3mm} \\
  1668.865\tablenotemark{b} &  0.570 &  $-$39.2  \hspace*{3mm} &    108.5   \hspace*{3mm} \\
  1669.867\tablenotemark{b} &  0.746 &  $-$93.6  \hspace*{3mm} &    246.8   \hspace*{3mm} \\
  1670.858\tablenotemark{b} &  0.920 &  $-$45.4  \hspace*{3mm} &    120.0   \hspace*{3mm} \\
  1671.862\tablenotemark{b} &  0.097 &     53.7  \hspace*{3mm} & $-$137.6   \hspace*{3mm} \\
  1672.851\tablenotemark{b} &  0.271 &     92.6  \hspace*{3mm} & $-$247.0   \hspace*{3mm} \\
  2037.839\tablenotemark{b} &  0.513 &   $-$4.6  \hspace*{3mm} &   $-$1.3   \hspace*{3mm} \\
  2039.826\tablenotemark{b} &  0.862 &  $-$71.9  \hspace*{3mm} &    188.3   \hspace*{3mm} \\
  2040.837\tablenotemark{b} &  0.040 &     22.4  \hspace*{3mm} &  $-$55.4   \hspace*{3mm} \\
  2335.830\tablenotemark{b} &  0.962 &  $-$17.6  \hspace*{3mm} &    108.7   \hspace*{3mm} \\
  2336.809\tablenotemark{b} &  0.134 &     71.8  \hspace*{3mm} & $-$180.3   \hspace*{3mm} \\
  2337.786\tablenotemark{b} &  0.306 &     87.4  \hspace*{3mm} & $-$233.1   \hspace*{3mm} \\
  2338.770\tablenotemark{b} &  0.479 &      8.8  \hspace*{3mm} &  \nodata   \hspace*{3mm} \\
  2339.781\tablenotemark{b} &  0.657 &  $-$77.0  \hspace*{3mm} &    204.0   \hspace*{3mm} \\
  2381.748\tablenotemark{b} &  0.044 &     22.9  \hspace*{3mm} &  $-$68.1   \hspace*{3mm} \\
  2382.714\tablenotemark{b} &  0.214 &     91.2  \hspace*{3mm} & $-$241.5   \hspace*{3mm} \\
  2383.712\tablenotemark{b} &  0.390 &     58.9  \hspace*{3mm} & $-$159.7   \hspace*{3mm} \\
  3130.693\tablenotemark{c} &  0.865 &  $-$70.1  \hspace*{3mm} &    185.1   \hspace*{3mm} \\
  3131.731\tablenotemark{c} &  0.048 &     26.8  \hspace*{3mm} &  $-$70.0   \hspace*{3mm} \\
  3132.765\tablenotemark{c} &  0.230 &     93.7  \hspace*{3mm} & $-$247.8   \hspace*{3mm} \\
  3133.800\tablenotemark{c} &  0.412 &     46.7  \hspace*{3mm} & $-$127.4   \hspace*{3mm} \\
  3134.675\tablenotemark{c} &  0.566 &  $-$38.9  \hspace*{3mm} &     97.5   \hspace*{3mm} \\
  3135.718\tablenotemark{c} &  0.750 &  $-$93.9  \hspace*{3mm} &    245.4   \hspace*{3mm} \\
\tableline
\end{tabular}\\
\tablecomments{Instrumental setups: $^{a}$ ESO CAT + CES; $^{b}$ ESO~1.5m + FEROS; $^{c}$ ESO/MPG~2.2m + FEROS.
RVs are quoted in the zero velocity reference frame. For future conversion, we adopted the systemic velocities $\overline{\gamma_1}=-21.5$\,\kms\ and  $\overline{\gamma_2}=-23.1$\,\kms\ (see \S\ref{ssect: orbit}).}
\end{table}

\clearpage

\begin{table*}
\caption{Journal of the \xmm\ observations.
\label{tab: journal} }

\centering          

\begin{tabular}{c c c c  c c c c}
\tableline
\tableline
\vspace*{-3mm}\\
 JD    & $\phi_\mathrm{He~I}$ & \multicolumn{3}{c}{Effective duration (ks)} &  \multicolumn{3}{c}{Count rates ($10^{-3}$ \cnts)}\\
$-2\,450\,000$ &   & \mos1 & \mos2 & \pn  & \mos1 & \mos2 & \pn  \\ 
\tableline
 2158.214 &  0.700  & 33.1 & 33.2 & 30.6 & $10.1\pm0.7$ & $11.4\pm0.7$                & $36.0\pm1.4$ \\
 2158.931 &  0.826  & 19.8 & 19.8 & 16.5 & $10.7\pm1.0$ & $10.4\pm0.9$                & $40.2\pm2.1$ \\
 2159.796 &  0.978  & 33.7 & 33.9 & 30.1 & $11.0\pm0.7$ & $ \hspace*{1.5mm}9.8\pm0.7$ & $36.7\pm1.5$ \\
 2160.925 &  0.177  & 26.0 & 24.3 & 11.7 & $12.0\pm0.9$ & $11.3\pm0.9$                & $46.0\pm2.6$ \\
 2161.774 &  0.326  & 30.9 & 31.0 & 28.4 & $11.9\pm0.8$ & $12.5\pm0.8$                & $38.6\pm1.5$ \\
 2162.726 &  0.494  & 32.9 & 32.8 & 30.3 & $ 9.8\pm0.7$ & $10.6\pm0.7$                & $36.0\pm1.4$ \\
\tableline
\end{tabular}
\end{table*}


\clearpage
\thispagestyle{empty}
{\rotate
\begin{table*}
\caption{ \label{tab: orbit_ne} Orbital solutions deduced from different RV data sets (i.e.~associated to different lines).}
\begin{tabular}{c c c c c c c c c  }
\tableline
\tableline
\vspace*{-3mm}\\
Lines       & $P$      &$s_y/s_x$ &  $T_\mathrm{ic}$~\tablenotemark{a}      &  $K_1$          & $K_2$          &   $\gamma_1$        &$\gamma_2$        &r.m.s.         \\
            &     (d)  &          &  HJD$-$2\,450\,000      &        (\kms)   &        (\kms)  &              (\kms) &           (\kms) &       (\kms)  \\
\tableline
\hea\,\lam4471  &  5.68151   & 4.7   & 3199.634 $\pm$ 0.004 &  94.2 $\pm$ 0.5 & 245.5 $\pm$ 1.2 & $-$22.6 $\pm$ 0.4 & $-$23.0 $\pm$ 0.5 &  2.0 \\
\hea\,\lam4922  &  5.68149   & 4.2   & 3199.627 $\pm$ 0.006 &  94.5 $\pm$ 0.8 & 245.7 $\pm$ 2.0 & $-$20.7 $\pm$ 0.7 & $-$22.1 $\pm$ 0.9 &  3.2 \\
\hea\,\lam5016  &  5.68153   & 3.7   & 3199.640 $\pm$ 0.003 &  93.3 $\pm$ 0.4 & 247.9 $\pm$ 1.0 & $-$22.5 $\pm$ 0.3 & $-$26.4 $\pm$ 0.3 &  1.9 \\
\hea\,\lam5048  &  5.68150   & 2.1   & 3199.639 $\pm$ 0.004 &  92.4 $\pm$ 0.7 & 249.7 $\pm$ 1.7 & $-$22.4 $\pm$ 0.6 & $-$25.4 $\pm$ 0.7 &  3.2 \\
\hea\,\lam5876  &  5.68148   & 3.6   & 3199.630 $\pm$ 0.002 &  93.2 $\pm$ 0.3 & 246.7 $\pm$ 0.7 & $-$20.5 $\pm$ 0.3 & $-$20.8 $\pm$ 0.3 &  1.3 \\
\hea\,\lam7065  &  5.68147   & 3.4   & 3199.641 $\pm$ 0.003 &  94.6 $\pm$ 0.4 & 242.4 $\pm$ 1.0 & $-$19.6 $\pm$ 0.5 & $-$20.9 $\pm$ 0.4 &  1.8 \\
\tableline							 
\heb\,\lam4686  &  5.68154   &  --   & 3199.640 $\pm$ 0.002 &  94.6 $\pm$ 0.5 &  --             & $-$16.5 $\pm$ 0.2 &   --            &  1.5 \\
\oc\,\lam5592   &  5.68150   &  --   & 3199.630 $\pm$ 0.002 &  94.7 $\pm$ 0.3 &  --             & $-$24.4 $\pm$ 0.2 &   --            &  1.1 \\
\cc\,\lam5696   &  5.68155   &  --   & 3199.643 $\pm$ 0.002 &  93.8 $\pm$ 0.8 &  --             & $-$34.3 $\pm$ 0.2 &   --            &  3.1 \\
\tableline 
\end{tabular}\\
\tablenotetext{a}{ $T_\mathrm{ic}$ is the time of inferior conjunction, the primary being in front of the secondary.}
\end{table*}
}
\clearpage


\begin{table}
\caption{Orbital and physical parameters of \cpd. \label{tab: orbit}}

\begin{tabular}{c c c c}
\tableline
\tableline
                       & Prim.            & \hea\ lines          & Lit.            \\
\tableline
$P$ (d)                & 5.681514         &   5.681504        & 5.681534         \\      
                         & $\pm$  3.91\,10$^{-4}$          &   $\pm$  3.18\,10$^{-4}$        & $\pm$  4.7\,10$^{-5}$           \\     
$s_y/s_x$              &  \nodata         &   3.3                & \nodata \vspace*{2mm} \\
$M_1/M_2$              &  \nodata         &   2.640 $\pm$ 0.012  & \nodata \vspace*{2mm} \\
$T_\mathrm{ic}$ (HJD    & 3199.632         &    3199.635          & 3199.640         \\
$-$2\,450\,000)        &\hspace*{2mm}$\pm$ 0.002 & \hspace*{1mm} $\pm$ 0.003&\hspace*{1mm} $\pm$ 0.002\vspace*{2mm}\\
$K_1$\ (\kms)          &  93.6 $\pm$ 0.3  &    93.7 $\pm$ 0.3    & 91.9 $\pm$ 0.2   \\
$K_2$\ (\kms)          &  \nodata         &   247.4 $\pm$ 0.9    & \nodata               \\
$v_{0,1}$\ (\kms)     & $-0.3  \pm$ 0.3  &   0.0   $\pm$ 0.2    & 0.0 $\pm$ 0.1  \\
$v_{0,2}$\ (\kms)     & \nodata          &  $-$0.3 $\pm$ 0.3    & \nodata               \\
$a_1 \sin i$ (\rsol)   & 10.51 $\pm$ 0.06 &   10.52 $\pm$ 0.06   & 10.31$\pm$ 0.05  \\
$a_2 \sin i$ (\rsol)   &  \nodata         &   27.76 $\pm$ 0.10   & \nodata \vspace*{2mm}  \\
$M_1 \sin^3 i$ (\msol) &  \nodata         &   16.94 $\pm$ 0.15   & \nodata               \\
$M_2 \sin^3 i$ (\msol) &  \nodata         &    6.42 $\pm$ 0.05   & \nodata   \\
$f_\mathrm{mass}$(\msol)& 0.483 $\pm$ 0.004 &  \nodata             & 0.457 $\pm$ 0.003  \vspace*{2mm}     \\

r.m.s. (\kms)          & 1.5              & 1.9                  &  5.7             \\
\tableline
\end{tabular}
\tablecomments{Note that the present fits are done in the zero velocity frame. Hence the systemic velocities in this frame (noted $v_{0,i}$) are expected to be zero (see \S\ref{ssect: orbit} and \S\ref{ssect: lit} for values in the heliocentric frame).}
\end{table}


\clearpage

\begin{table}
\caption{Average equivalent widths ($W_\lambda$) of selected \cpd\ spectral lines}
\label{tab: lines_list}
\begin{tabular}{c c c}
\tableline
\tableline
\vspace*{-3mm} & \\
 Line      & $W_{\lambda,1}$ & $W_{\lambda,2}$\\
             & (m\AA) &  (m\AA)\\
\tableline
\hea\,\lam4121 & 161: &  21:   \\
\hea\,\lam4144 & $188\pm15$ &  $61\pm12$   \\
\hea\,\lam4388 & $273\pm21$ &  $83\pm25$   \\
\hea\,\lam4471 & $635\pm35$ &  100:   \\
\hea\,\lam5016 & $255\pm5$ & $20\pm4$  \\
\hea\,\lam5876 & $945\pm11$ & $58\pm2$  \\
\heb\,\lam4542 & $382\pm28$ &  \nodata   \\
\heb\,\lam4686 & $590\pm20$ &  \nodata   \\
 \cb\,\lam4267 & $39\pm3$    &  $11\pm2$   \\
 \cc\,\lam5696 & $104\pm8$    & \nodata   \\
 \nb\,\lam4630 & $105\pm5$ &   7:    \\
 \ob\,\lam4070 & blend &   9:    \\
 \oc\,\lam5592 & $251\pm7$ & \nodata   \\
\sic\,\lam4553 &  $71\pm4$ &   $7\pm2$:    \\
\sic\,\lam4575 &  $29\pm5$:&  7:    \\
\sid\,\lam4089 & $354\pm16$ &  \nodata    \\
\mgb\,\lam4481 & $95\pm9$ &  $20\pm3$    \\

\tableline
\end{tabular}\\
\tablecomments{The $W_\lambda$ are referred to the total flux of the binary system.}
\end{table}

\clearpage

\thispagestyle{empty}
{\rotate
\begin{table*}
\caption{ \epic\ spectra best fit parameters and observed X-ray fluxes. \label{tab: Xspec}} 

\begin{tabular}{c c c c c c c c c c c }
\tableline
\tableline
\vspace*{-3mm}\\
$\phi_\mathrm{He~I}$ &   k$T_1$ & $norm_1$               & $N_\mathrm{H, 2}$ \tablenotemark{a} & k$T_2$ & $norm_2$              & $\chi^2_{\nu}$ (d.o.f.) & $f_\mathrm{X}$ & $f_\mathrm{X,S}$ & $f_\mathrm{X,M}$  & $f_\mathrm{X,H}$ \\
                     & (keV)    & ($10^{-5}$\,cm$^{-5}$) & ($10^{22}$\,cm$^{-2}$)   & (keV)  & ($10^{-5}$\,cm$^{-5}$)&                         & \multicolumn{4}{c}{($10^{-14}$\,erg\,cm$^{-2}$\,s$^{-1}$)}       \\
\tableline
\vspace*{-3mm}\\
 0.700   &  $0.26^{0.28}_{0.23}$ & $10.2^{11.6}_{8.8}$ & $0.54^{0.81}_{0.36}$ & $0.73^{0.86}_{0.61}$ & $ 5.8^{ 7.5}_{4.2}$ & 1.06 (87) & 7.3 & 4.6 & 2.4 & 0.2  \vspace*{1mm}\\ 
 0.826   &  $0.19^{0.27}_{0.14}$ & $12.4^{21.4}_{8.4}$ & $0.26^{0.71}_{0.01}$ & $0.50^{0.66}_{0.37}$ & $ 9.5^{19.8}_{5.8}$ & 0.82 (56) & 8.0 & 5.3 & 2.7 & 0.1  \vspace*{1mm}\\
 0.978   &  $0.28^{0.30}_{0.25}$ & $ 9.6^{11.0}_{8.5}$ & $0.58^{0.96}_{0.34}$ & $0.74^{0.83}_{0.63}$ & $ 5.6^{ 8.4}_{3.6}$ & 0.80 (94) & 7.5 & 4.8 & 2.5 & 0.2  \vspace*{1mm}\\ 
 0.177   &  $0.25^{0.30}_{0.21}$ & $10.8^{12.5}_{8.3}$ & $0.80^{1.17}_{0.52}$ & $0.67^{2.58}_{0.50}$ & $12.3^{23.8}_{3.1}$ & 1.23 (54) & 8.2 & 4.7 & 3.2 & 0.3  \vspace*{1mm}\\
 0.326   &  $0.26^{0.31}_{0.19}$ & $10.4^{11.9}_{8.9}$ & $0.47^{0.91}_{0.16}$ & $0.63^{1.12}_{0.52}$ & $ 7.1^{10.4}_{2.2}$ & 1.14 (89) & 7.9 & 5.0 & 2.7 & 0.2  \vspace*{1mm}\\ 
 0.494   &  $0.28^{0.31}_{0.24}$ & $8.6^{9.9 }_{7.3} $ & $0.41^{0.74}_{0.21}$ & $0.71^{0.87}_{0.62}$ & $ 4.8^{ 6.8}_{3.1}$ & 1.25 (87) & 7.0 & 4.5 & 2.4 & 0.2  \vspace*{1mm}\\ 
\tableline
\vspace*{-3mm}\\	       
  Merged  & $0.28^{0.29}_{0.26}$ & $9.9^{10.5}_{9.3} $ & $0.54^{0.67}_{0.42}$ & $0.80^{1.01}_{0.74}$ & $ 5.1^{ 6.0}_{4.2}$ & 1.37 (315)& 7.7 & 4.8 & 2.6 & 0.2  \vspace*{1mm}\\ 
\tableline
\end{tabular}
\tablenotetext{a}{ As indicated by its subscript,  the local equivalent Hydrogen column $N_\mathrm{H, 2}$  only applies to the second temperature component k$T_2$ of the model.}
\end{table*}
}

\clearpage

\begin{table}
\centering
\caption{Absorption corrected fluxes and X-ray luminosity of \cpd. \label{tab: xlum}}

\begin{tabular}{c c c c c c} 
\tableline
\tableline
$\phi_\mathrm{He~I}$ &  $f_\mathrm{X}^\mathrm{un.}$ & $f_\mathrm{X,S}^\mathrm{un.}$ & $f_\mathrm{X,M}^\mathrm{un.}$  & $f_\mathrm{X,H}^\mathrm{un.}$ & $\log L_\mathrm{X}$ \vspace*{1mm}\\
       & \multicolumn{4}{c}{($10^{-14}$\,erg\,cm$^{-2}$\,s$^{-1}$)} & (\ergs)\\
\tableline
\multicolumn{6}{c}{2-T models} \\
\tableline
 0.700  & 21.3 & 17.6 & 3.5 & 0.2 & 31.83 \\
 0.826  & 25.7 & 21.6 & 3.9 & 0.1 & 31.92 \\
 0.978  & 21.0 & 17.1 & 3.6 & 0.2 & 31.83 \\
 0.177  & 23.0 & 18.3 & 4.4 & 0.3 & 31.87 \\ 
 0.326  & 22.8 & 18.8 & 3.9 & 0.2 & 31.86 \\ 
 0.494  & 19.6 & 16.0 & 3.4 & 0.2 & 31.80 \\ 
  Merged  &  21.5 & 17.5 & 3.8 & 0.2 & 31.84 \\ 
\tableline
\multicolumn{6}{c}{3-T model} \\
\tableline
  Merged  &  22.4 & 18.0 & 3.8 & 0.6 & 31.86 \\ 
\tableline
\multicolumn{6}{c}{2-T + PL model} \\
\tableline
  Merged  & 22.4 & 18.0 & 3.8 & 0.6 & 31.86 \\
\tableline
\end{tabular}
\end{table}

\clearpage




\clearpage

\clearpage
   \begin{figure}
   \centering
   \includegraphics[width=.9\columnwidth]{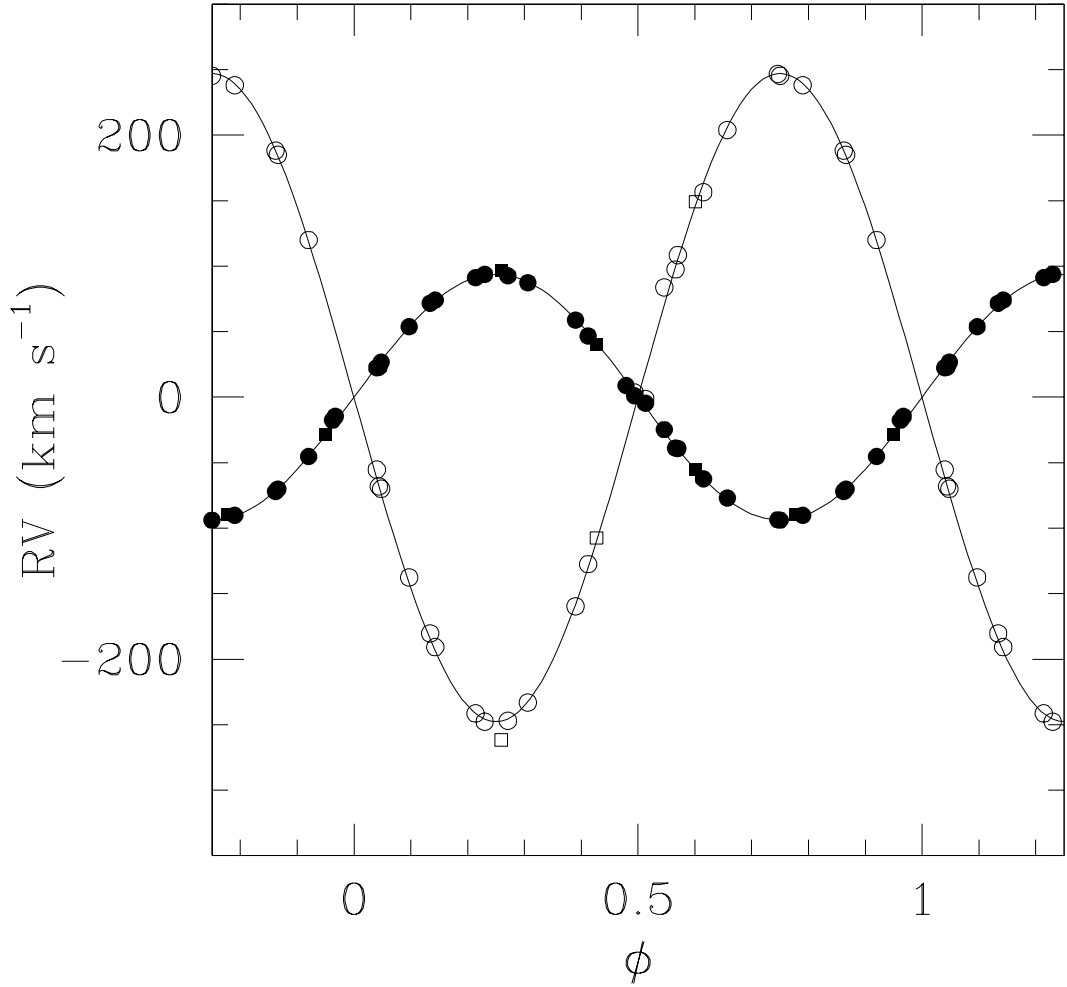}
      \figcaption{\cpd\ RV curves  corresponding to the circular \hea\ solution of Table \ref{tab: orbit}. The RV measurements listed in Table \ref{tab: opt_diary} are also displayed. Filled and open symbols correspond to the primary and secondary components respectively. The squares give the CES data while the circles indicate the FEROS measurements.
        \label{fig: sb2}       }
        
   \end{figure}



\clearpage


\begin{thebibliography}{}

\bibitem[\protect\citeauthoryear{{Arnaud}}{{Arnaud}}{1996}]{arn96}
{Arnaud}, K.~A. 1996, in ASP Conf. Ser. 101, Astronomical Data
  Analysis Software and Systems V, eds. G. Jacoby, \& J. Barnes, 17


\bibitem[\protect\citeauthoryear{{Conti}}{{Conti}}{1973a}]{Con73_nlte}
{Conti}, P.~S. 1973a, \apj, 179, 161

\bibitem[\protect\citeauthoryear{{Conti}}{{Conti}}{1973b}]{Con73_teff}
{Conti}, P.~S. 1973b, \apj, 179, 181

\bibitem[\protect\citeauthoryear{{Conti} \& {Alschuler}}{{Conti} \&
  {Alschuler}}{1971}]{CA71}
{Conti}, P.~S., \&  {Alschuler}, W.~R. 1971, \apj, 170, 325

\bibitem[\protect\citeauthoryear{{Conti}, {Leep} \& {Lorre}}{{Conti}
  et~al.}{1977}]{CLL77}
{Conti}, P.~S.,  {Leep}, E.~M., \& {Lorre}, J.~J. 1977, \apj, 214, 759

\bibitem[\protect\citeauthoryear{{Didelon}}{{Didelon}}{1982}]{Did82}
{Didelon}, P. 1982, \aaps, 50, 199

\bibitem[\protect\citeauthoryear{{Gies}}{2003}]{Gie03}
{Gies}, D.~R. 2003,  in IAU Symp. 212, A Massive Star Odyssey: From Main Sequence to Supernova, eds. K. van der Hucht, A. Herrero, \& C. Esteban, 91

\bibitem[\protect\citeauthoryear{{Gosset}, {Royer}, {Rauw}, {Manfroid} \&
  {Vreux}}{{Gosset} et~al.}{2001}]{GRR01}
{Gosset}, E.,  {Royer}, P.,  {Rauw}, G.,  {Manfroid}, J.,   \&  {Vreux}, J.-M. 2001,
  \mnras, 327, 435

\bibitem[\protect\citeauthoryear{{Heck}, {Manfroid} \& {Mersch}}{{Heck}
  et~al.}{1985}]{HMM85}
{Heck}, A.,  {Manfroid}, J., \&  {Mersch}, G. 1985, \aaps, 59, 63

\bibitem[\protect\citeauthoryear{{Hill}, {Crawford} \& {Barnes}}{{Hill}
  et~al.}{1974}]{HCB74}
{Hill}, G.,  {Crawford}, D.~L., \& {Barnes}, J.~V. 1974, \aj, 79, 1271

\bibitem[\protect\citeauthoryear{{Howarth} \& {Prinja}}{{Howarth} \&
  {Prinja}}{1989}]{HP89}
{Howarth}, I.~D., \&  {Prinja}, R.~K. 1989, \apjs, 69, 527

\bibitem[\protect\citeauthoryear{{Humphreys} \& {McElroy}}{{Humphreys} \&
  {McElroy}}{1984}]{HM84}
{Humphreys}, R.~M., \&  {McElroy}, D.~B. 1984, \apj, 284, 565

\bibitem[\protect\citeauthoryear{{Kaastra}}{{Kaastra}}{1992}]{Ka92}
{Kaastra}, J. 1992, {An X-Ray Spectral Code for Optically Thin Plasmas}.
(Internal SRON-Leiden Report, updated version 2.0)

\bibitem[\protect\citeauthoryear{{Lafler} \& {Kinman}}{{Lafler} \&
  {Kinman}}{1965}]{LK65}
{Lafler}, J., \&  {Kinman}, T.~D. 1965, \apjs, 11, 216

\bibitem[\protect\citeauthoryear{{Lanz} \& {Hubeny}}{{Lanz} \& {Hubeny}}{2003}]{LaH03}
{Lanz}, T., \& {Hubeny}, I. 2003, \apjs, 146, 417 (Erratum: ApJS, 147, 225)

\bibitem[\protect\citeauthoryear{{Levato} \& {Morrell}}{{Levato} \&
  {Morrell}}{1983}]{LM83}
{Levato}, H., \&  {Morrell}, N. 1983, \aplett, 23, 183

\bibitem[\protect\citeauthoryear{{Martins}, {Schaerer}\&{Hillier}}{{Martins} et~al.}{2005}]{Martins}
{Martins}, F., {Schaerer}, D., \& {Hillier}, D.J. 2005, \aap, 436, 1049

\bibitem[\protect\citeauthoryear{{Mathys}}{{Mathys}}{1988}]{Mat88}
{Mathys}, G. 1988, \aaps, 76, 427

\bibitem[\protect\citeauthoryear{{Mathys}}{{Mathys}}{1989}]{Mat89}
{Mathys}, G. 1989, \aaps, 81, 237

\bibitem[\protect\citeauthoryear{{Mermilliod}}{{Mermilliod}}{1988}]{Me88}
{Mermilliod}, J.-C. 1988, Bulletin d'Information du Centre de Donn\'ees
  Stellaires (CDS), 35, 77

\bibitem[\protect\citeauthoryear{{Mermilliod}}{{Mermilliod}}{1992}]{Me92}
{Mermilliod}, J.-C. 1992, Bulletin d'Information du Centre de Donn\'ees
  Stellaires (CDS), 40, 115

\bibitem[\protect\citeauthoryear{{Mewe}, {Gronenschild} \& {van den
  Oord}}{{Mewe} et~al.}{1985}]{MGvdO85}
{Mewe}, R.,  {Gronenschild}, E.~H.~B.~M.,  \&   {van den Oord}, G.~H.~J. 1985,
  \aaps, 62, 197

\bibitem[\protect\citeauthoryear{{Moore}}{{Moore}}{1959}]{Moore59}
{Moore}, C.~E. 1959, {A multiplet table of astrophysical interest. Part 1}.
NBS Technical Note, Washington: US Department of Commerce, 1959, Rev.~edition

\bibitem[\protect\citeauthoryear{{Perry}, {Hill}, {Younger} \&
  {Barnes}}{{Perry} et~al.}{1990}]{PHYB90}
{Perry}, C.~L.,  {Hill}, G.,  {Younger}, P.~F.,   \&  {Barnes}, J.~V. 1990, \aaps,
  86, 415

\bibitem[\protect\citeauthoryear{{Raboud}, {Cramer} \& {Bernasconi}}{{Raboud}
  et~al.}{1997}]{RCB97}
{Raboud}, D.,  {Cramer}, N.,  \&   {Bernasconi}, P.~A. 1997, \aap, 325, 167

\bibitem[\protect\citeauthoryear{{Sana}, {Hensberge}, {Rauw} \&
  {Gosset}}{{Sana} et~al.}{2003}]{SHRG03}
{Sana}, H.,  {Hensberge}, H.,  {Rauw}, G.,  \&   {Gosset}, E. 2003, \aap, 405, 1063

\bibitem[\protect\citeauthoryear{{Sana}, {Stevens}, {Gosset}, {Rauw} \&
  {Vreux}}{{Sana} et~al.}{2004}]{SSG04}
{Sana}, H.,  {Stevens}, I.~R.,  {Gosset}, E.,  {Rauw}, G.,   \&  {Vreux}, J.-M. 2004,
  \mnras, 350, 809

\bibitem[\protect\citeauthoryear{{Sana}, {Antokhina}, {Royer}, {Manfroid},
  {Gosset}, {Rauw} \& {Vreux}}{{Sana} et~al.}{2005}]{SAR05}
{Sana}, H.,  {Antokhina}, E.,  {Royer}, P.,  {Manfroid}, J.,  {Gosset}, E.,  {Rauw}
  G.,  \&   {Vreux}, J.-M. 2005, \aap, 441, 213

\bibitem[\protect\citeauthoryear{{Sana}, {Gosset}, {Rauw}, {Sung} \&
  {Vreux}}{{Sana} et~al.}{2006a}]{SGR06}
{Sana}, H.,  {Gosset}, E.,  {Rauw}, G.,  {Sung}, H., \&    {Vreux}, J.-M. 2006a, \aap, 454, 1047 

\bibitem[\protect\citeauthoryear{{Sana}, {Gosset} \& {Rauw}}{{Sana}
  et~al.}{2006b}]{SGR06_219}
{Sana}, H.,  {Gosset}, E.,  \&   {Rauw}, G. 2006b, \mnras, 371, 67

\bibitem[\protect\citeauthoryear{{Sana}, {Rauw}, {Naz\'e}, {Gosset}, \& {Vreux}}{{Sana}
  et~al.}{2006c}]{SRN06}
{Sana}, H., {Rauw}, G., {Naz\'e}, Y., {Gosset}, E. \& {Vreux}, J.-M. 2006c, \mnras, 372, 661

\bibitem[\protect\citeauthoryear{{Schaller}, {Schaerer}, {Meynet} \&
  {Maeder}}{{Schaller} et~al.}{1992}]{SSM92}
{Schaller}, G.,  {Schaerer}, D.,  {Meynet}, G., \&    {Maeder}, A. 1992, \aaps, 96,
  269

\bibitem[\protect\citeauthoryear{{Schmidt-Kaler}}{{Schmidt-Kaler}}{1982}]{SK82}
{Schmidt-Kaler}, T. 1982, {Physical Parameters of the Stars}.
Vol.~2b of {Landolt-B\"ornstein, Numerical Data and Functional Relationships in
  Science and Technology, New Series, Group VI}, Springer-Verlag, Berlin

\bibitem[\protect\citeauthoryear{{Seggewiss}}{{Seggewiss}}{1968}]{Se68a}
{Seggewiss}, W. 1968, Veroeffentlichungen des Astronomischen Institute der
  Universitaet Bonn, 79

\bibitem[\protect\citeauthoryear{{Stickland} \& {Lloyd}}{{Stickland} \&
  {Lloyd}}{2001}]{SL01}
{Stickland}, D.~J., \&  {Lloyd}, C. 2001, The Observatory, 121, 1

\bibitem[\protect\citeauthoryear{{Struve}}{{Struve}}{1944}]{Str44}
{Struve}, O. 1944, \apj, 100, 189

\bibitem[\protect\citeauthoryear{{Sung}, {Bessell} \& {Lee}}{{Sung}
  et~al.}{1998}]{SBL98}
{Sung}, H.,  {Bessell}, M.~S., \& {Lee}, S. 1998, \aj, 115, 734

\bibitem[\protect\citeauthoryear{{Underhill}}{{Underhill}}{1994}]{Und94}
{Underhill}, A.~B. 1994, \apj, 420, 869

\bibitem[\protect\citeauthoryear{{Usov}}{{Usov}}{1992}]{Usov}
{Usov}, V.~V. 1992, \apj, 389, 635

\bibitem[\protect\citeauthoryear{{Vink}, {de Koter} \& {Lamers}}{{Vink}
  et~al.}{2001}]{VdKL01}
{Vink}, J.~S.,  {de Koter}, A.,  \&   {Lamers}, H.~J.~G.~L.~M. 2001, \aap, 369, 574

\bibitem[\protect\citeauthoryear{{Walborn}}{{Walborn}}{1973}]{Wal73}
{Walborn}, N.~R.  1973, \aj, 78, 1067

\bibitem[\protect\citeauthoryear{{Walborn} \& {Fitzpatrick}}{{Walborn} \&
  {Fitzpatrick}}{1990}]{WF90}
{Walborn}, N.~R., \&  {Fitzpatrick}, E.~L.  1990, \pasp, 102, 379


\end{thebibliography}
\end{document}